\documentclass[aps,twocolumn,prb,eqsecnum,showpacs]{revtex4}
\usepackage[dvips]{graphicx}
\begin{document}
\draft
\preprint{21 June 2008}
\title[Halogen-bridged diplatinum complexes]
      {Photoproduction of spin and charge carriers in
       halogen-bridged binuclear platinum chain complexes}
\author{Shoji Yamamoto and Jun Ohara}
\address{Department of Physics, Hokkaido University,
         Sapporo 060-0810, Japan}
%\date{Received \hspace{4cm}}
\date{21 June 2008}
\begin{abstract}
Nonlinear lattice relaxation of photoexcited diplatinum-halide chain
compounds is theoretically investigated within a one-dimensional extended
Peierls-Hubbard model.
We first illuminate the whole relaxation scenario in terms of
variational wave functions and then visualize each relaxation channel
numerically integrating the Schr\"odinger equation.
High-energy excitations above the electron-hole continuum tend to relax
into polarons, while excitons pumped within the optical gap, unless
luminescent, turn into solitonic states nonradiatively.
Neutral and charged solitons coexist as stable photoproducts, which has
never been observed in conventional platinum-halide chains, and they
are highly resonant on the occasion of their birth and geminate
recombination.
\end{abstract}
% 42.65.-k: Nonlinear optics
% 42.65.Tg: Optical solitons; nonlinear guided waves
% 71.10.Hf: Non-Fermi-liquid ground states, electron phase diagrams and
%           phase transitions in model systems
% 71.23.An: Theories and models; localized states
% 71.35.-y: Excitons and related phenomena 
% 71.35.Aa: Frenkel excitons and self-trapped excitons 
% 71.38.-k: Polarons and electron-phonon interactions
% 71.45.Lr: Charge-density-wave systems
% 71.55.-i: Impurity and defect levels
% 75.40.Mg: Numerical simulation studies
% 78.20.Bh: Theory, models, and numerical simulation
% 78.20.Ci: Optical constants (including refractive index,
%           complex dielectric constant, absorption, reflection and
%           transmission coefficients)
% 78.30.-j: Infrared and Raman spectra
% 78.47.+p: Time-resolved optical spectroscopies and other ultrafast
%           optical measurements in condensed matter 
% 78.55.-m: Photoluminescence, properties and materials 
\maketitle

\section{Introduction}

   Halogen ($X$)-bridged binuclear transition-metal ($M$) linear-chain
complexes \cite{C4604,C409,B444,B2815} have stimulated a renewed
interest in the $M\!X$ class of materials. \cite{G6408,W6435}
Metal binucleation induces an unpaired electron per metal-dimer unit even
in valence-trapped states, contrasting with the Peierls-gapped state of
$M^{2+}$ and $M^{4+}$ in conventional $M\!X$ chains.
The $M\,d_{z^2}$-$M\,d_{z^2}$ direct overlap effectively reduces the
on-site Coulomb repulsion due to its $d_{\sigma^*}$ character,
remarkably enhancing the electron itinerancy.
The $M\!M\!X$ electronic state is thus activated and full of variety.
\cite{Y125124,K2163}
Diplatinum-halide varieties,
 $A_4$[Pt$_2$(pop)$_4X$]$\cdot$$m$H$_2$O
($X=\mbox{Cl},\mbox{Br},\mbox{I}$;
 $A=\mbox{NH}_4,\mbox{Na},\mbox{K},\cdots$;
 $\mbox{pop}=\mbox{diphosphonate}=\mbox{P}_2\mbox{O}_5\mbox{H}_2$)
\cite{C4604,C409}
structurally resemble conventional $M\!X$ compounds and exhibit the most
likely charge density wave (CDW) with the halogen-sublattice dimerized:
\cite{B1155,K40}
$-X^{-}\!\cdots\mbox{Pt}^{2+}\mbox{Pt}^{2+}\!\cdots
  X^{-}\!-\mbox{Pt}^{3+}\mbox{Pt}^{3+}\!-X^{-}\!\cdots$.
However, the ground state is quite sensitive to bridging halogens
and counter ions. \cite{Y2321,Y13}
In the iodo complexes the valence arrangement is highly tunable with
pressure application and optical irradiation.
\cite{S1405,K18682,Y140102,Y1489,M046401,Y075113}
Their analog without any counter ion,
 Pt$_2$(dta)$_4$I
($\mbox{dta}=\mbox{dithioacetate}
 =\mbox{CH}_3\mbox{CS}_2$) \cite{B444} is of metallic
conduction above room temperature. \cite{K1931,C5552}
With decreasing temperature, it undergoes successive phase transitions
\cite{K10068,Y1198,I115110,I387} and ends up with a distinct
Peierls-distorted state of alternating charge polarization (ACP):
\cite{B4562,W6676}
$\cdots\mbox{I}^{-}\!\cdots\mbox{Pt}^{2+}\mbox{Pt}^{3+}\!-\mbox{I}^{-}
 \!-\!\mbox{Pt}^{3+}\mbox{Pt}^{2+}\!\cdots\mbox{I}^{-}\!\cdots$,
where the metal sublattice, which is free from any hydrogen bonding, is
dimerized.
The methyl group in the dta ligand can be replaced by longer alkyl chains
\cite{M11179,M2767} so as to enhance the one dimensionality.

   Photoexcited $M\!X$ complexes are more and more interesting.
By analogy with topological excitations in the trans isomer of
polyacetylene, several authors \cite{I137,O250,B339} had an idea of
various defect states existing in Pt$X$ chains.
Solitonic and polaronic excitations \cite{G6408,W6435,M5758,M5763,G10566}
were particularly investigated in relation to experimental findings.
Such charge and/or spin carriers may be less massive and more mobile in
Pt$_2X$ chains with much smaller band gaps. \cite{Y189,Y165113}
There indeed appear new vibrational features in the resonant Raman
spectra of K$_4$[Pt$_2$(pop)$_4$Cl]$\cdot$H$_2$O, which can be ascribed to
polaronic defects due to the deficiency of counter ions. \cite{C723}
Electron-spin-resonance (ESR) measurements on
Pt$_2$(C$_5$H$_{11}$CS$_2$)$_4$I
reveal thermally activated mobile spins attributable to neutral solitons.
\cite{T2169}
Photoexcitation is a promising approach to producing these valence
anomalies in a systematic fashion.
Thus motivated, we analyze the electron-lattice dynamics of
photoexcited Pt$_2X$ chains.
We first reveal the whole relaxation scenario in terms of variational
wave functions and then visualize each story as a solution of
Schr\"odinger equation.
Neutral solitons and/or polarons were observed in
[Pt(en)$_2$Cl](ClO$_4$)$_2$
($\mbox{en}=\mbox{ethylenediamine}=\mbox{C}_2\mbox{H}_8\mbox{N}_2$),
\cite{S3066,K1789,D49}
whereas charged solitons and polarons in
[Pt(en)$_2$I](ClO$_4$)$_2$. \cite{O2248,O6330}
However, neutral and charged solitons do not seem able to coexist as
photoproducts in any $M\!X$ chain. \cite{M5758,I1088}
There may be a wider variety of relaxation channels in varied $M\!M\!X$
chains. \cite{O045122}

\begin{figure}[b]
\centering
\includegraphics[width=70mm]{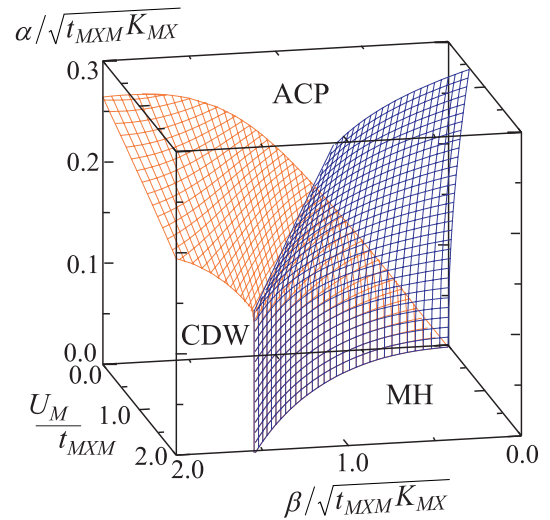}
%\vspace*{-3mm}
\caption{%(Colour online)
         Hartree-Fock calculation of a ground-state phase diagram on the
         $\alpha$-$\beta$-$U_M$ cube for $M\!M\!X$ chains.}
\label{F:PhD}
\end{figure}

\section{Modelling}

   We describe Pt$_2X$ chains by the $\frac{3}{4}$-filled single-band
Peierls-Hubbard Hamiltonian
\begin{eqnarray}
   &&\!\!\!\!
   {\cal H}
   =-\sum_{n,s}
     \bigl[t_{M\!X\!M}-\alpha(l_{n+1}+r_{n})\bigr]
     \bigl(a_{n+1,s}^\dagger b_{n,s}
   \nonumber \\
   &&\ \ \mbox{}
    +b_{n,s}^\dagger a_{n+1,s}\bigr)
    -t_{M\!M}\sum_{n,s}
     \bigl(b_{n,s}^\dagger a_{n,s}+a_{n,s}^\dagger b_{n,s}\bigr)
   \qquad
   \nonumber \\
   &&\ \ \mbox{}
    -\beta\sum_{n,s}
     (l_{n}n_{n,s}+r_{n}m_{n,s})
    +\frac{K_{M\!X}}{2}\sum_{n}
     \bigl(l_{n}^2
    +r_{n}^2\bigr)
   \nonumber \\
   &&\ \ \mbox{}
    +\frac{M_X}{2}\sum_n{\dot u}_n^2
    +M_M\sum_n{\dot v}_n^2
   \nonumber \\
   &&\ \ \mbox{}
    +U_{M}\sum_{n}
     (n_{n,\uparrow}n_{n,\downarrow}+m_{n,\uparrow}m_{n,\downarrow})
   \nonumber \\
   &&\ \ \mbox{}
    +\sum_{n,s,s'}
     (V_{M\!M}n_{n,s}m_{n,s'}+V_{M\!X\!M}n_{n+1,s}m_{n,s'}),
   \label{E:H}
\end{eqnarray}
where
$n_{n,s}=a_{n,s}^\dagger a_{n,s}$ and
$m_{n,s}=b_{n,s}^\dagger b_{n,s}$ with
$a_{n,s}^\dagger$ and $b_{n,s}^\dagger$ creating an electron with spin
$s=\uparrow,\downarrow\equiv\pm$ on the $\mbox{Pt}\,d_{z^2}$ orbitals in
the $n$th diplatinum unit.
$t_{M\!M}$ and $t_{M\!X\!M}$ describe the intradimer electron transfer
and the interdimer electron supertransfer, respectively, and are set for
$t_{M\!M}=2t_{M\!X\!M}$.
$\alpha$ and $\beta$ characterize electron-lattice interactions of the
Peierls and Holstein types, respectively.
$l_{n}=v_n-u_{n-1}$ and $r_{n}=u_n-v_n$ with $u_n$ and $v_n$ being,
respectively, the chain-direction displacements of the $n$th halogen ion
and diplatinum cluster from their equilibrium positions.
Deformation of every diplatinum cluster is negligible.
$K_{M\!X}$ is the platinum-halogen spring constant, while
$M_X$ and $2M_M$ are the masses of a halogen atom and a diplatinum
complex, respectively.
Any calculation is carried out in a chain of a hundred or more Pt$_2X$
units under the periodic boundary condition, where no significant size
effect survives.

   We show in Fig. \ref{F:PhD} a ground-state phase diagram of the
Hamiltonian (\ref{E:H}) within the Hartree-Fock (HF) approximation.
The site-diagonal Holstein-type electron-lattice coupling stabilizes the
CDW state against the Mott-Hubbard (MH) insulating state, while the
site-off-diagonal Peierls-type one contributes towards realizing the ACP
state.
We consider both pop and dta families of Pt$_2X$ complexes, setting
$(\alpha,\beta)/\sqrt{t_{M\!X\!M}K_{M\!X}}$ equal to
$(0.0,1.2)$ and $(0.3,0.8)$, respectively.
Typical Pt$X$ chains lie in the intermediate-correlation regime
\cite{W6435,M5758} of $V_{M\!X\!M}\ll U_M\alt t_{M\!X\!M}$
and platinum binucleation should effectively reduce the on-site repulsion.
Thus we fix the Pt$_2X$ Coulomb parameters at
$(U_M,V_{M\!M},V_{M\!X\!M})/t_{M\!X\!M}=(0.5,0.25,0.15)$.
The $d$-$p$ hybridization and the resultant interdimer supertransfer
significantly depend on the bridging halogens.
The Pt-I transfer integral is indeed a few times as large as the Pt-Cl
one. \cite{C723,S1659}
The present parametrization well features Pt$_2X$ chains, but
the Coulomb parameters, when scaled, may fundamentally vary with $X$.

\section{Variational Calculation}

   Photoinduced charge-transfer excitations first spread over the chain
and then transform themselves into local defects.
Such self-localization processes can be visualized through the calculation
of adiabatic potential energy surfaces in a variational manner.
\cite{M5758,M5763,I1088,O045122}
Solitonic relaxation channels are describable with a trial wave function
\begin{eqnarray}
   &&
   l_{n}
  =\sigma(-1)^n\lambda
   \Biggl[
    1+\kappa
      \biggl({\rm tanh}\frac{|n|-d/2}{\xi}-1\biggr)
   \Biggr],
   \nonumber \\
   &&
   r_{n}
  =(-1)^n\lambda
   \Biggl[
    1+\kappa
      \biggl({\rm tanh}\frac{|n+\delta|-d/2}{\xi}-1\biggr)
   \Biggr],\qquad
   \label{E:WFS}
\end{eqnarray}
whereas polaronic ones with
\begin{eqnarray}
   &&
   l_{n}
  =\sigma(-1)^n\lambda
   \Biggl[
    1+\kappa
      \biggl({\rm tanh}\biggl|\frac{|n|-d/2}{\xi}\biggr|-1\biggr)
   \Biggr],
   \nonumber \\
   &&
   r_{n}
  =(-1)^n\lambda
   \Biggl[
    1+\kappa
      \biggl({\rm tanh}\biggl|\frac{|n+\delta|-d/2}{\xi}\biggr|-1\biggr)
   \Biggr],\qquad\ 
   \label{E:WFP}
\end{eqnarray}
where the variational parameters $\kappa$, $\xi$ and $\delta$ are
determined at every interdefect distance $d$ given so as to minimize the
energy of the lowest-lying excited state, setting $\lambda$ equal to the
uniform halogen-ion displacement relative to neighbouring diplatinum
clusters in the CDW ($\sigma=1$) or ACP ($\sigma=-1$) ground state.
Once the ground state is photoexcited into the Frank-Condon state, which
still sits at $\kappa=0$, the uniform bond alternation begins to be
locally deformed.
Increasing $\kappa$ with $d$ remaining to be zero depicts the
self-stabilization of a charge-transfer exciton.
The fully stabilized, that is to say, self-trapped exciton (STE) may have
paths to a pair of soliton and antisoliton and that of polarons.
Such defect pairs may be directly generated from free electron-hole pairs,
which are higher-lying excited states, with $\kappa$ and $d$ being tuned
simultaneously.
$\xi$ determines the width of a defect.
$\delta$ is just a minor parameter, bringing the defect center any other
types of valence oscillation besides the ground-state one. \cite{Y165113}
\begin{figure*}
\centering
\includegraphics[width=170mm]{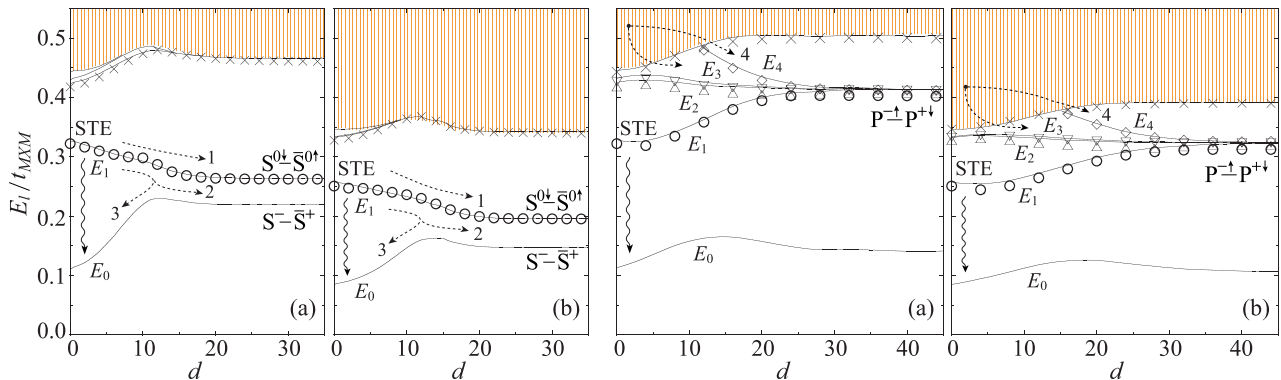}
\vspace*{-3mm}
\caption{%(Colour online)
         The solitonic (the left two panels) and polaronic (the right two
         panels) relaxation channels of photoexcited Pt$_2X$ chains whose
         ground states are of the CDW (a) and ACP (b) types, which are
         relevant to
         $A_4$[Pt$_2$(pop)$_4X$]$\cdot$$m$H$_2$O and
         Pt$_2$(dta)$_4$I, respectively.
         $E_l^{\rm HF}$ is plotted with solid lines, while $E_l^{\rm CI}$
         with symbols.
         A fully stabilized exciton, unless decays by luminescence, may be
         dissociated into a neutral (1) or charged (2)
         soliton (S)-antisoliton ($\bar{\mbox{S}}$) pair.
         Any potential soliton pair is so fragile as to recombine to
         disappear (3).
         Polaron (P) pairs may be generated by high-energy excitations
         reaching the electron-hole continuum.}
\label{F:ElPt2X}
\end{figure*}

   The lattice configuration is adiabatically determined within the HF
scheme.
The ground-state bond-alternation amplitude $\lambda$ minimizes
$\langle E_0|{\cal H}|E_0\rangle\equiv E_0$ in a chain of constant length,
where
\begin{equation}
   |E_0\rangle
  =\prod_{\epsilon_{\mu,s}\leq\epsilon_{\rm F}}
   c_{\mu,\uparrow}^\dagger c_{\mu,\downarrow}^\dagger|0\rangle,
\end{equation}
with $|0\rangle$ denoting the true electron vacuum, $\epsilon_{\rm F}$
indicating the Fermi energy, and $c_{\mu,s}^\dagger$, generally given as
$\sum_n
(\vartheta_{\mu,n,s}a_{n,s}^\dagger+\varphi_{\mu,n,s}b_{n,s}^\dagger)$,
creating an electron of spin $s$ in the HF eigenstate of eigenvalue
$\epsilon_{\mu,s}$.
The excited-state parameters $\kappa$, $\xi$ and $\delta$ minimize
$_{\rm HF}\langle E_1|{\cal H}|E_1\rangle_{\rm HF}$, where
\begin{equation}
   |E_l\rangle_{\rm HF}
  =c_{\nu,s}^\dagger c_{\mu,s}|E_0\rangle,
\end{equation}
provided $\epsilon_{\mu,s}\leq\epsilon_{\rm F}<\epsilon_{\nu,s}$.
For the first excited state $l=1$, $\mu$ and $\nu$ are set equal to the
highest-lying occupied and lowest-lying vacant levels, respectively.
When we take particular interest in the $l$th energy surface,
$_{\rm HF}\langle E_l|{\cal H}|E_l\rangle_{\rm HF}$ may be minimized
instead.
However, the whole energy scheme is not so sensitive to the variational
target as to be qualitatively changeable.
In any case the HF $l$th excited-state energy $E_l^{\rm HF}$ is
trivially written as $E_0+\epsilon_{\nu,s}-\epsilon_{\mu,s}$.

   We further consider configuration interactions (CIs) between the HF
excited states in an attempt to describe Coulomb correlations with more
precision.
Excited states of the single-particle-hole CI type are expressed as
\begin{equation}
   |E_l\rangle_{\rm CI}
   =\sum_{\epsilon_{\mu,s}\leq\epsilon_{\rm F}<\epsilon_{\nu,s}}
    f(\mu,\nu,s;l)c_{\nu,s}^\dagger c_{\mu,s}|E_0\rangle,
\end{equation}
where $f(\mu,\nu,s;l)$ diagonalizes the original Hamiltonian (\ref{E:H}).
The refined $l$th excited-state energy $E_l^{\rm CI}$ is given by
$_{\rm CI}\langle E_l|{\cal H}|E_l\rangle_{\rm CI}$.

   Figures \ref{F:ElPt2X} presents the thus-calculated adiabatic potential
energies as functions of $d$, where the ground-state energy is set for
zero.
Even though fully stabilized excitons are most likely to decay directly
into the ground state, whether radiatively or nonradiatively, we take
particular interest in their relaxing further into solitonic states.
No energy barrier between an STE and any $\mbox{S}-\bar{\mbox{S}}$ pair, 
which is not the case with Pt$X$ chains, \cite{M5758,I1088} allows STEs
to decay rather fast.
Solitons have localized wave functions and their overlap rapidly decreases
with increasing $d$.
Instantaneous charge transport between far distant S and $\bar{\mbox{S}}$
is hardly feasible.
Therefore, tunnelling between the energy surfaces $E_1$ and $E_0$ occurs at
a moderate separation.
Charged soliton pairs may be less reachable than neutral ones, because
assistant lattice fluctuations of odd parity \cite{I1088} are necessary to
their occurrence.
In any case, once a pair of soliton and antisoliton is generated at a
certain distance, say, $d\agt 20$, they must be long-lived owing to an
energy barrier to the ground state, especially on an ACP background.
On the other hand, there is no possibility of an STE changing into a
polaron pair.
Polaronic states are available from the electrons pumped above the optical
gap.
Polaron pairs disappear into STEs, which are either luminescent or ready
for further relaxation into solitonic states
(see Figs. \ref{F:PPonCDW} and \ref{F:PPonACP} later on).

   Indeed there are Coulomb-correlation-induced configuration interactions
especially in the small-$d$ region, but the correction
$E_l^{\rm CI}-E_l^{\rm HF}$ is not so significant in Pt$_2X$ chains of
moderate Coulomb correlations.
It is not the case with Ni$_2$I chains of $U_M\gg t_{M\!X\!M}$.
\cite{O1571}
Since the most important variational prediction of neutral and charged
solitons coexisting as stable photoproducts remains unchanged with the CI
refinement of the energy surfaces, we reasonably employ the time-dependent
HF method in order to verify this scenario.

\section{Real Time Dynamics}

   We trace charge-transfer excitations as functions of real time solving
the Schr\"odinger equation
\begin{equation}
   {\rm i}\hbar{\dot{\mit\Psi}}_{\mu,s}(t)
  ={\cal H}_s^{\rm HF}(t){\mit\Psi}_{\mu,s}(t),
   \label{E:Schrodinger}
\end{equation}
where ${\cal H}_s^{\rm HF}(t)$ and ${\mit\Psi}_{\mu,s}(t)$, the spin-$s$
sectors of the HF Hamiltonian and wave function, are given by a square
matrix and a column vector of degree $2N$, respectively.
Defining the wave vector as
\begin{equation}
   {\mit\Psi}_{\mu,s}(t)
  =\left[
    \begin{array}{c}
     \vartheta_{\mu,1,s}(t) \\
     \varphi_{\mu,1,s}(t)   \\
     \vdots                 \\
     \vartheta_{\mu,N,s}(t) \\
     \varphi_{\mu,N,s}(t)
    \end{array}
   \right],
\end{equation}
and employing a unitary transformation

\noindent
\begin{equation}
   U=\left[
      \begin{array}{ccccc}
       0 & 1 & 0 & \cdots & 0 \\
       0 & 0 & 1 & \ddots & \vdots \\
       \vdots & \ddots & \ddots & \ddots & 0 \\
       0 & \cdots & 0 & 0 & 1 \\
       1 & 0 & \cdots & 0 & 0
      \end{array}
     \right],
\end{equation}
\quad

\noindent
we express the Hamiltonian as
\begin{eqnarray}
   &&\!\!\!\!\!\!\!\!\!\!\!\!
   {\cal H}_s^{\rm HF}(t)
  ={\rm diag}
   \left[
    {\cal P}_{1,s}(t),\cdots,{\cal P}_{N,s}(t)
   \right]
   \nonumber \\
   &&\quad\mbox{}
  +U^\dagger
   {\rm diag}
   \left[
    {\cal Q}_{1,s}(t),\cdots,{\cal Q}_{N,s}(t)
   \right]
   U,
   \label{E:CBD}
\end{eqnarray}
where the $2\times 2$ matrices
\begin{widetext}
\begin{eqnarray}
   &&
   {\cal P}_{n,s}(t)
  =\left[
    \begin{array}{cc}
      \begin{array}{c}
      -{\displaystyle\frac{\beta}{2}}[v_{n}(t)-u_{n-1}(t)]
      +{\displaystyle\frac{U_{M}}{2}}A_{n,-s}(t) \\
      \mbox{}
      +{\displaystyle\frac{V_{M\!M}}{2}}B_n(t)
      +{\displaystyle\frac{V_{M\!X\!M}}{2}}B_{n-1}(t)
      \end{array}
    &
      -t_{M\!M}-V_{M\!M}P_{n,s}(t) 
    \\
      -t_{M\!M}-V_{M\!M}P_{n,s}^*(t)
    &
      \begin{array}{c}
      -{\displaystyle\frac{\beta}{2}}[u_{n}(t)-v_{n}(t)]
      +{\displaystyle\frac{U_{M}}{2}}B_{n,-s}(t) \\
      \mbox{}
      +{\displaystyle\frac{V_{M\!M}}{2}}A_n(t)
      +{\displaystyle\frac{V_{M\!X\!M}}{2}}A_{n+1}(t)
      \end{array}
    \end{array}
   \right],
   \nonumber \\
   &&
   {\cal Q}_{n,s}(t)
  =\left[
    \begin{array}{cc}
      \begin{array}{c}
      -{\displaystyle\frac{\beta}{2}}[u_{n}(t)-v_{n}(t)]
      +{\displaystyle\frac{U_{M}}{2}}B_{n,-s}(t) \\
      \mbox{}
      +{\displaystyle\frac{V_{M\!M}}{2}}A_n(t)
      +{\displaystyle\frac{V_{M\!X\!M}}{2}}A_{n+1}(t)
      \end{array}
    &
      \begin{array}{c}
      -t_{M\!X\!M}+\alpha[v_{n+1}(t)-v_{n}(t)] \\
      \mbox{}
      -V_{M\!X\!M}Q_{n,s}(t) 
      \end{array}
    \\
      \begin{array}{c}
      -t_{M\!X\!M}+\alpha[v_{n+1}(t)-v_{n}(t)] \\
      \mbox{}
      -V_{M\!X\!M}Q_{n,s}^*(t)
      \end{array}
    &
      \begin{array}{c}
      -{\displaystyle\frac{\beta}{2}}[v_{n+1}(t)-u_{n}(t)]
      +{\displaystyle\frac{U_{M}}{2}}A_{n+1,-s}(t) \\
      +{\displaystyle\frac{V_{M\!M}}{2}}B_{n+1}(t)
      +{\displaystyle\frac{V_{M\!X\!M}}{2}}B_{n}(t)
      \end{array}
    \end{array}
   \right],
\end{eqnarray}
\end{widetext}
are self-consistently calculated through
\begin{eqnarray}
   &&\!\!\!\!\!\!\!\!\!\!\!\!\!\!\!\!\!\!
   A_{n}(t)=\sum_{s}A_{n,s}(t)
  =\sum_{s}\mathop{{\sum}'}_{\mu}
   \bigl|\vartheta_{\mu,n,s}(t)\bigr|^2,
   \nonumber \\
   &&\!\!\!\!\!\!\!\!\!\!\!\!\!\!\!\!\!\!
   B_{n}(t)=\sum_{s}B_{n,s}(t)
  =\sum_{s}\mathop{{\sum}'}_{\mu}
   \bigl|\varphi_{\mu,n,s}(t)\bigr|^2,
   \nonumber \\
   &&\!\!\!\!\!\!\!\!\!\!\!\!\!\!\!\!\!\!
   P_{n}(t)=\sum_{s}P_{n,s}(t)
  =\sum_{s}\mathop{{\sum}'}_{\mu}
   \vartheta_{\mu,n,s}^*(t)\varphi_{\mu,n,s}(t),
   \nonumber \\
   &&\!\!\!\!\!\!\!\!\!\!\!\!\!\!\!\!\!\!
   Q_{n}(t)=\sum_{s}Q_{n,s}(t)
  =\sum_{s}\mathop{{\sum}'}_{\mu}
   \varphi_{\mu,n,s}^*(t)\vartheta_{\mu,n+1,s}(t),
\end{eqnarray}
with $\sum^\prime$ denoting the summation over the initially occupied
levels.
Now the HF wave vector $\{\vartheta_{\mu,n,s},\varphi_{\mu,n,s}\}$ and the
lattice configuration $\{u_n,v_n\}$ are both time dependent.
Discretizing the time variable as $t_j=j{\mit\Delta}t$ $(j=0,1,2,\cdots)$
with the interval ${\mit\Delta}t$ much smaller than
$\omega_{\rm eff}^{-1}\equiv\sqrt{M_{M}M_{X}/(2M_{M}+M_{X})K_{M\!X}}$,
we schematically integrate eq. (\ref{E:Schrodinger}):
\begin{equation}
   {\mit\Psi}_{\mu,s}(t_{j+1})
  ={\hat T}{\rm exp}
   \Bigl[
   -\frac{{\rm i}}{\hbar}\int_{t_j}^{t_j+{\mit\Delta}t}
    {\cal H}_s^{\rm HF}(t){\rm d}t
   \Bigr]
   {\mit\Psi}_{\mu,s}(t_{j}),
   \label{E:Psi}
\end{equation}
where ${\hat T}$ denotes the time ordering.
Employing a general decomposition theory of ordered exponentials,
\cite{S387} together with the expression (\ref{E:CBD}), we can carry out
the time evolution of eq. (\ref{E:Psi}) without numerically
diagonalizing ${\cal H}_s^{\rm HF}(t)$ at every time step, \cite{K703}
which serves to accelerate the calculation.
We adopt a fractal decomposition up to the third order of
${\mit\Delta}t$. \cite{S161,T177}

   The lattice dynamics is governed by Newton's equation of motion,
\begin{eqnarray}
   &&\!\!\!\!\!\!\!\!
   M_{X}{\ddot u}_{n}(t)
  =K_{M\!X}[v_{n+1}(t)-2u_{n}(t)+v_{n}(t)]
   \nonumber \\
   &&\quad\mbox{}
  -\beta[A_{n+1}(t)-B_n(t)]\equiv F_{n}(t),
   \nonumber \\
   &&\!\!\!\!\!\!\!\!
   2M_{M}{\ddot v}_{n}(t)
  =K_{M\!X}[u_{n}(t)-2v_{n}(t)+u_{n-1}(t)]
   \nonumber \\
   &&\quad\mbox{}
  +\alpha[Q_{n}(t)+Q_{n}^*(t)-Q_{n-1}(t)-Q_{n-1}^*(t)]
   \nonumber \\
   &&\quad\mbox{}
  +\beta[A_{n}(t)-B_{n}(t)]\equiv G_{n}(t).
\end{eqnarray}
With the discrete time variable, we evolve the lattice configuration as
\begin{eqnarray}
   &&\!\!\!\!\!\!\!\!
   u_{n}(t_{j+1})
  =u_{n}(t_{j})+{\dot u}_{n}(t_{j}){\mit\Delta}t,
   \nonumber \\
   &&\!\!\!\!\!\!\!\!
   {\dot u}_{n}(t_{j+1})
  ={\dot u}_{n}(t_{j})+\frac{F_{n}(t_{j})}{M_X}{\mit\Delta}t,
   \nonumber \\
   &&\!\!\!\!\!\!\!\!
   v_{n}(t_{j+1})
  =v_{n}(t_{j})+{\dot v}_{n}(t_{j}){\mit\Delta}t,
   \nonumber \\
   &&\!\!\!\!\!\!\!\!
   {\dot v}_{n}(t_{j+1})
  ={\dot v}_{n}(t_{j})+\frac{G_{n}(t_{j})}{2M_M}{\mit\Delta}t.
\end{eqnarray}
We start any calculation from a stationary lattice of the CDW or ACP type:
\begin{eqnarray}
   &&\!\!\!\!\!\!\!\!
   v_{n}(t_0)-u_{n-1}(t_0)
  =\sigma(-1)^n\lambda+{\mit\Delta}l_{n}(t_0),
   \nonumber \\
   &&\!\!\!\!\!\!\!\!
   u_{n}(t_0)-v_{n}(t_0)
  =(-1)^n\lambda+{\mit\Delta}r_{n}(t_0),
   \label{E:ICcor}
   \\
   &&\!\!\!\!\!\!\!\!
   {\dot u}_{n}(t_0)={\dot v}_{n}(t_0)=0,
   \label{E:ICvel}
\end{eqnarray}
where ${\mit\Delta}l_{n}(t_0),{\mit\Delta}r_{n}(t_0)\ (\ll\lambda)$ are
introduced at random as thermal fluctuations.
\begin{figure*}
\centering
\includegraphics[width=160mm]{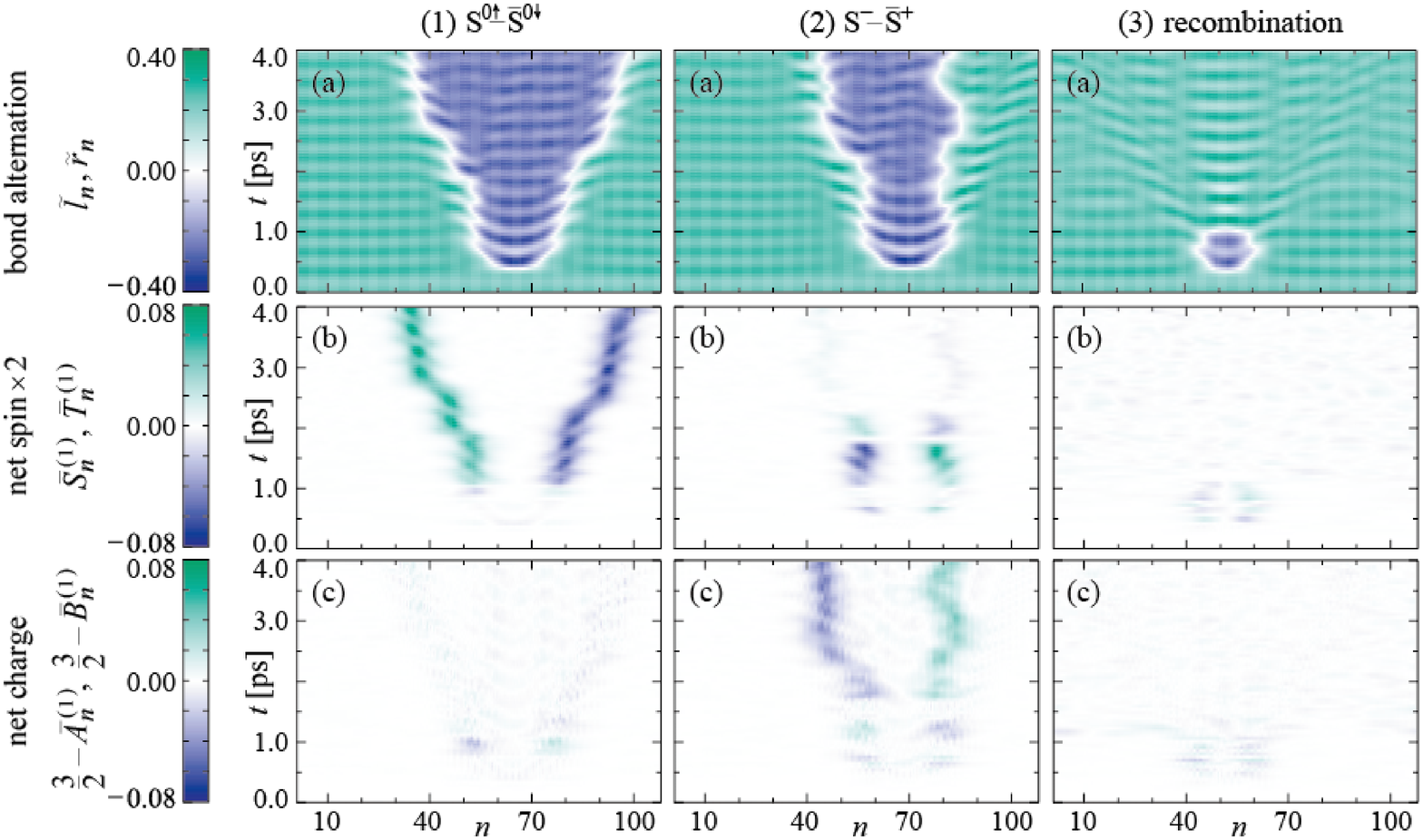}
\vspace*{-3mm}
\caption{%(Colour online)
         Contour plots of bond, spin and charge variables as functions
         of space and time for the first excited state of a Pt$_2X$ chain
         with a CDW ($\sigma=1$) background:
         (a) the bond order parameters
             ${\widetilde l}_{n}(t)
             =\sigma(-1)^n[-l_{n-1}(t)+2l_{n}(t)-l_{n+1}(t)]/4$,
             ${\widetilde r}_{n}(t)
             =(-1)^n[-r_{n-1}(t)+2r_{n}(t)-r_{n+1}(t)]/4$;
         (b) twice the net spin densities
             ${\bar S}_{n}^{(1)}(t)
             =\sum_{s}s{\bar A}_{n,s}^{(1)}(t)
             =\sum_{s}s[A_{n-1,s}^{(1)}(t)
                       +2A_{n,s}^{(1)}(t)
                       +A_{n+1,s}^{(1)}(t)]/4$,
             ${\bar T}_{n}^{(1)}(t)
             =\sum_{s}s{\bar B}_{n,s}^{(1)}(t)
             =\sum_{s}s[B_{n-1,s}^{(1)}(t)
                       +2B_{n,s}^{(1)}(t)
                       +B_{n+1,s}^{(1)}(t)]/4$;
         (c) the net electron densities
             ${\bar A}_{n}^{(1)}(t)
             =\sum_{s}{\bar A}_{n,s}^{(1)}(t)$,
             ${\bar B}_{n}^{(1)}(t)
             =\sum_{s}{\bar B}_{n,s}^{(1)}(t)$; where
         $l_{n}(t)=v_{n}(t)-u_{n-1}(t)$ and $r_{n}(t)=u_{n}(t)-v_{n}(t)$,
         while
         $A_{n,s}^{(l)}(t)=\sum_{\mu}
          f_{\mu,s}^{(l)}|\vartheta_{\mu,n,s}(t)|^2$ and
         $B_{n,s}^{(l)}(t)=\sum_{\mu}
          f_{\mu,s}^{(l)}|\varphi_{\mu,n,s}(t)|^2$ with
         $f_{\mu,s}^{(l)}$ being the time-independent distribution
         function of spin-$s$ electrons for the $l$th excited state.
         When $l=1$,
         $f_{\mu,s}^{(l)}=1$ for $\mu\leq 3N/2$ and
         $f_{\mu,s}^{(l)}=0$ for $\mu>3N/2$ but
         $f_{3N/2,\uparrow}^{(l)}=0$ and $f_{3N/2+1,\uparrow}^{(l)}=1$.}
\label{F:SSonCDW}
%\end{figure*}

%\begin{figure*}
\centering
\includegraphics[width=160mm]{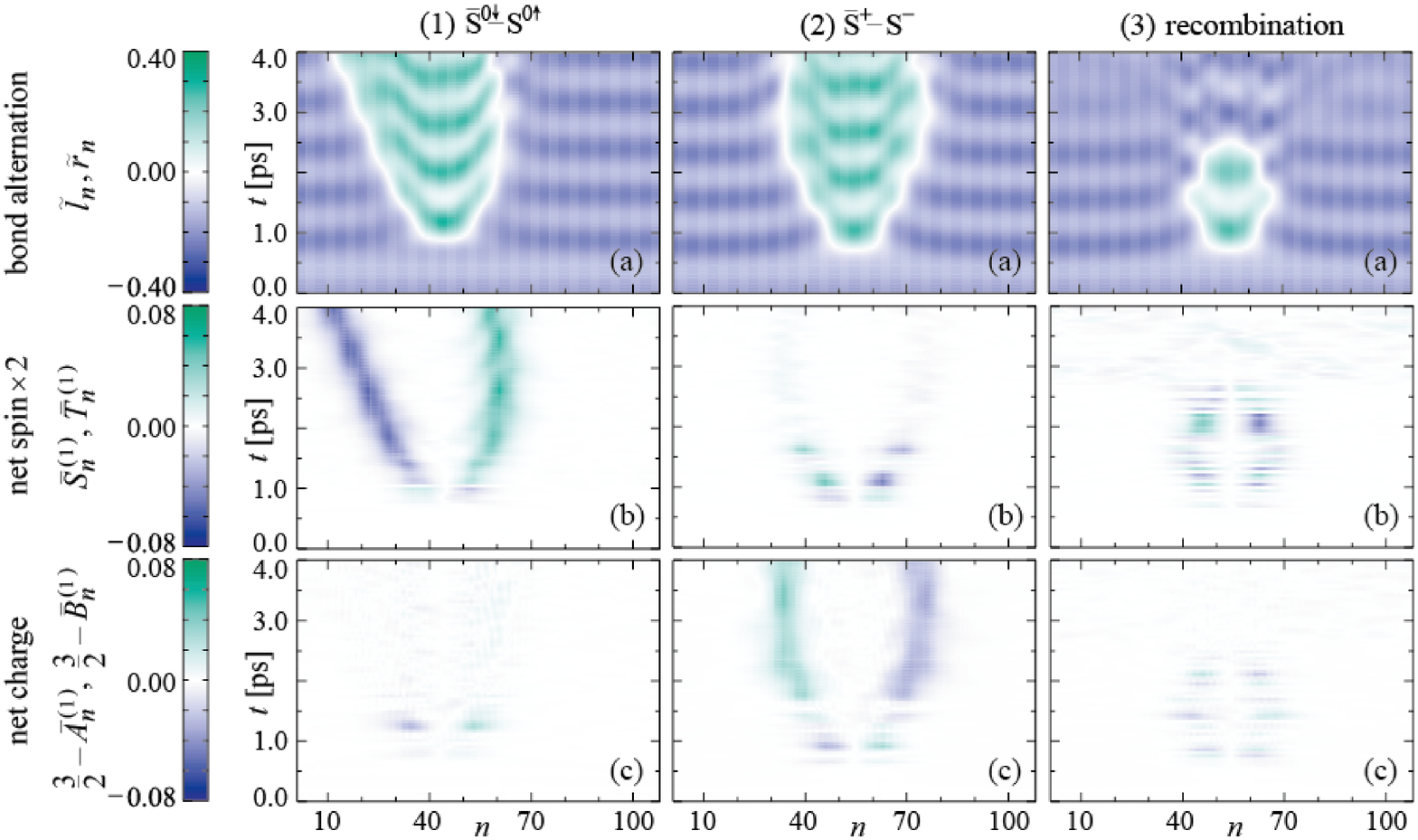}
\vspace*{-3mm}
\caption{%(Colour online)
         The same as Fig. \ref{F:SSonCDW} but with an ACP ($\sigma=-1$)
         background.}
\label{F:SSonACP}
\end{figure*}

\begin{figure*}
\centering
\includegraphics[width=118mm]{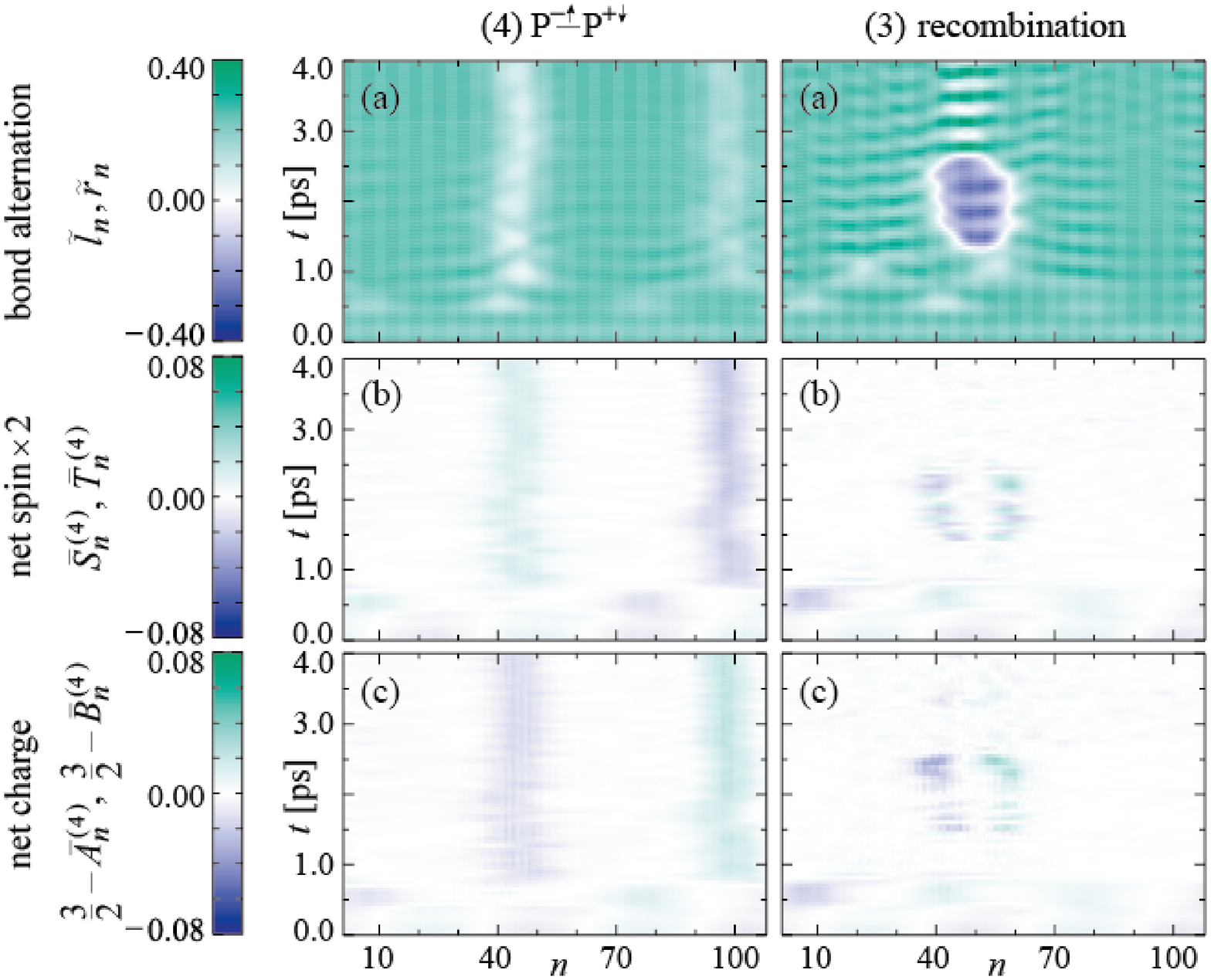}
\vspace*{-3mm}
\caption{%(Colour online)
         The same as Fig. \ref{F:SSonCDW} but for the fourth excited
         state.
         When $l=4$,
         $f_{\mu,s}^{(l)}=1$ for $\mu\leq 3N/2$ and
         $f_{\mu,s}^{(l)}=0$ for $\mu>3N/2$ but
         $f_{3N/2-1,\uparrow}^{(l)}=0$ and $f_{3N/2+2,\uparrow}^{(l)}=1$.}
\label{F:PPonCDW}
%\end{figure*}

%\begin{figure*}
\centering
\includegraphics[width=118mm]{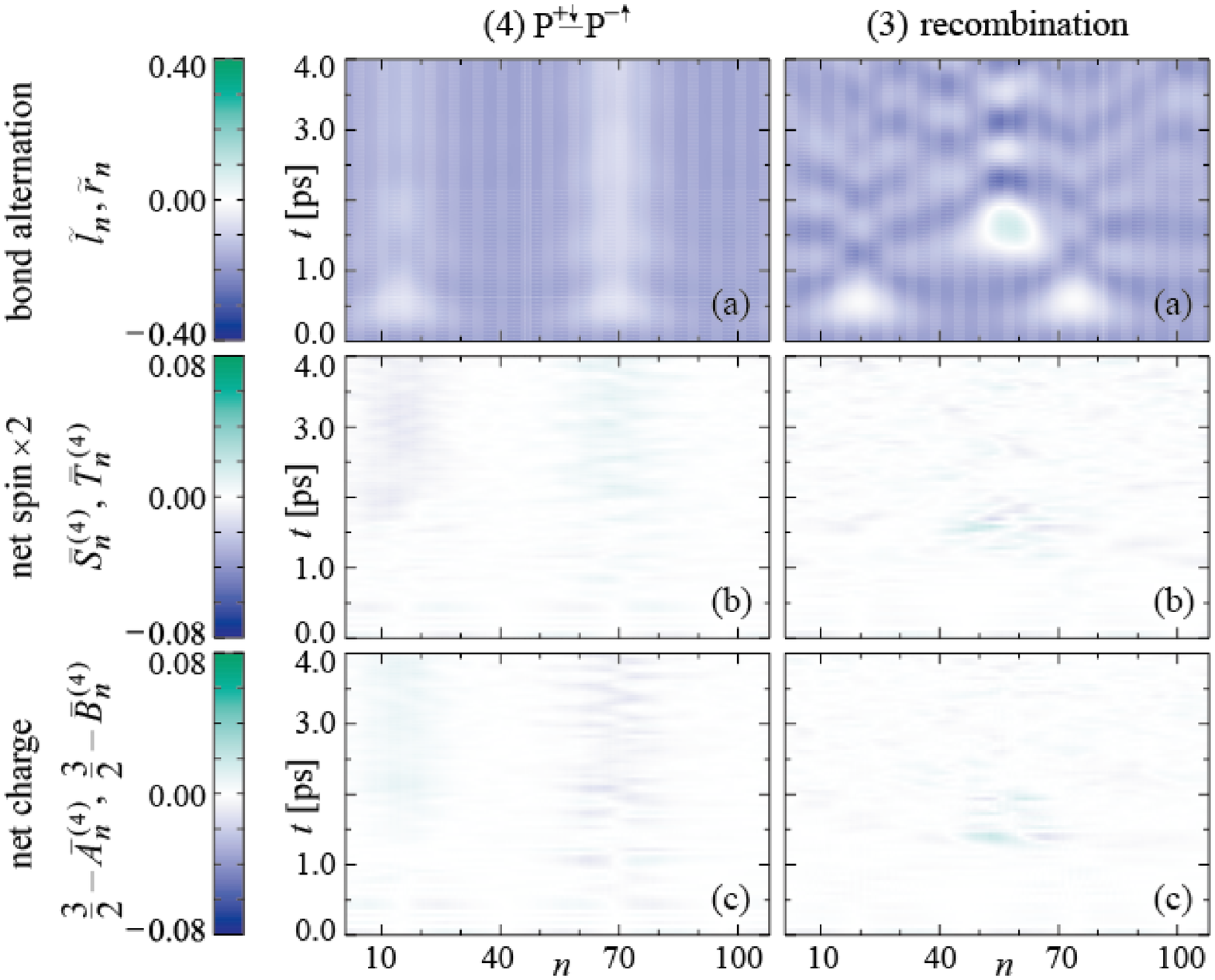}
\vspace*{-3mm}
\caption{%(Colour online)
         The same as Fig. \ref{F:PPonCDW} but with an ACP ($\sigma=-1$)
         background.}
\label{F:PPonACP}
\end{figure*}

   In an attempt to guide future experiments, we adopt a realistic set of
parameters:
$t_{M\!X\!M}=1.5\,\mbox{eV}$, $K_{M\!X}=8.0\,\mbox{eV}/\mbox{\AA}^2$,
$2M_{M}=13.4\times 10^5\,M_{\rm e}$, $M_{X}=2.3\times 10^5\,M_{\rm e}$;
\cite{Y189,Y165113} where $M_{\rm e}$ is the electron mass.
Numerical integration is reliable enough with
${\mit\Delta}t=10^{-3}\omega_{\rm eff}^{-1}=0.0255\,\mbox{fs}$.
The initial disorder [${\mit\Delta}l_{n}(t_0),{\mit\Delta}r_{n}(t_0)$] is
restricted to $10^{-4}\lambda$, which corresponds to low temperatures of
$10\,\mbox{K}
 \alt T\simeq 0.05\hbar\omega_{\rm eff}/k_{\rm B}
 \alt 20\,\mbox{K}$,
where both CDW and ACP ground states are fully stabilized.
\cite{S1405,K10068}

\begin{figure}[b]
\centering
\includegraphics[width=84mm]{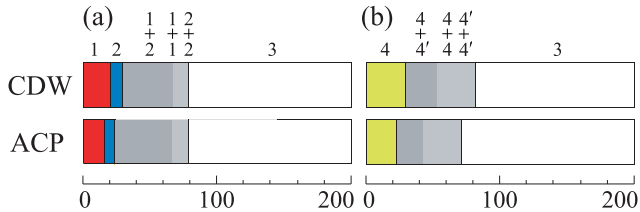}
\vspace*{-3mm}
\caption{%(Colour online)
         Characterization of photoproducts at the moment of four
         picoseconds after pumping up an electron to the first (a) and
         fourth (b) excited states with CDW and ACP backgrounds.
         Every case we observe two hundred samples changing the initial
         condition at random and encounter
         neutral soliton-antisoliton pairs ($1$),
         charged soliton-antisoliton pairs ($2$),
         linear combinations of them, which result in kink-antikink pairs
         with excess spin and charge densities of fraction ($1+2$) and
         those without any excess density of spin and charge ($1+1,2+2$),
         polaron pairs ($4,4'$),
         linear combination of them, which result in pairs of dips on the
         bond order parameters with either spin or charge excess density
         ($4+4'$) and those without any excess density of spin and charge
         ($4+4,4'+4'$), and
         no local defect detectable ($3$).
         The categorization is compactly summarized in
         Table \ref{T:photoproduct}.}
\label{F:photoproduct}
\end{figure}

   We first pump up an electron to the first excited state and observe the
following time evolution.
Typical examples are shown in Figures \ref{F:SSonCDW} and \ref{F:SSonACP}.
Photoproduction of neutral (1) and charged (2) soliton-antisoliton pairs
and their geminate recombination in the early stage (3) are clearly found
on both CDW and ACP backgrounds, supporting the analytical findings in
Fig. \ref{F:ElPt2X}.
In all the cases, a single and local defect is first nucleated on the
ground-state lattice configuration and further lattice relaxation induces
excess spin or charge density in the defect centers.
Although the photoexcitations are finalized into
$\mbox{S}^{0\uparrow}\!-\bar{\mbox{S}}^{0\downarrow}$ and
$\mbox{S}^{-}\!-\bar{\mbox{S}}^{+}$ pairs along the paths 1 and 2,
respectively, the early kinks on the path 2 convey net spins rather than
charges and then even wear charges and spins alternatively.
This is a strong evidence for tunnelling between the energy surfaces
$E_1$ and $E_0$ in Fig. \ref{F:ElPt2X}.
Kinks seem to be more frequently stabilized into spin solitons than into
charged ones at low temperatures (see Fig. \ref{F:photoproduct} later on).
The oscillation between neutral and charged kinks is generally observed on
the occasion of their geminate recombinations.

   The CDW and ACP backgrounds oscillate at regular intervals but with
different frequencies.
The CDW oscillation is mainly caused by the halogen sublattice, while
the ACP oscillation by the platinum sublattice.
The ratio of the former frequency to the latter one indeed reads
$\sqrt{2M_{M}/M_{X}}\simeq 2.4$.
Figures \ref{F:SSonCDW}(3) and \ref{F:SSonACP}(3) demonstrate in common
that a newborn soliton-antisoliton pair survives twice the oscillational
period and then recombine to disappear.
ACP solitons are consequently longer-lived than CDW solitons.

   Next we pump up an electron to the fourth excited state.
Typical examples are shown in Figures \ref{F:PPonCDW} and \ref{F:PPonACP}.
Now far distant defects in a pair, accompanied by excess spin and charge
densities all along, suddenly appear within much less than picoseconds,
which is again consistent with the analytical findings in
Fig. \ref{F:ElPt2X}.
Any defect is more delocalized on an ACP background than on a CDW
background and polarons are less localized than solitons in general.
The excess spin and charge densities accompanying the ACP polarons are
widely distributed.
All the findings accord with the preceding variational calculations.
When the interdefect distance $d$ is much larger than the defect width
$\xi$, the solitonic and polaronic wave functions (\ref{E:WFS}) and
(\ref{E:WFP}) are optimized at $\xi=3.45$ and $\xi=5.94$, respectively,
in the case of a CDW background, whereas they are most stable at
$\xi=4.25$ and $\xi=8.00$, respectively, in the case of an ACP background.
The smaller gap, the wider extent of any defect.
The adopted CDW and ACP ground states possess Peierls gaps of
$0.68\,\mbox{eV}$ and $0.53\,\mbox{eV}$, respectively.
The gap, unless too large, well scales the defect extent. \cite{Y165113}
The asymmetric lattice deformation of a pair of far distant polarons was
found in Pt$X$ chains as well. \cite{M5763}
A pair of polarons often recombine failing to stably separate away from
each other, where they disappear necessarily via an STE.
Unless the STE is luminescent, it nonradiatively relaxes further into a
solitonic state, as we have already observed in Figs. \ref{F:SSonCDW} and
\ref{F:SSonACP}.
Thus higher-energy excitations are longer-lived in general.
\begin{table}[b]
\caption{Classification of photoproducts.}
\begin{ruledtabular}
         \begin{tabular}{ccc}
         Character & Configuration & Label \\
         \noalign{\vskip 1mm}
         \hline
         \noalign{\vskip 1.5mm}

         neutral soliton &
         $\mbox{S}^{0\uparrow}\!-\bar{\mbox{S}}^{0\downarrow},
          \mbox{S}^{0\downarrow}\!-\bar{\mbox{S}}^{0\uparrow}$ &
         $1$ \\
         \noalign{\vskip 2mm}

         charged soliton &
         $\mbox{S}^{+}\!-\bar{\mbox{S}}^{-},
          \mbox{S}^{-}\!-\bar{\mbox{S}}^{+}$ &
         $2$ \\
         \noalign{\vskip 2mm}

         \begin{tabular}{c}
         kink with fractional \\
         spin and charge
         \end{tabular} &
         \begin{tabular}{c}
         $\mbox{K}^{\frac{+\uparrow}{2}}\!-
          \bar{\mbox{K}}^{\frac{-\downarrow}{2}},
          \mbox{K}^{\frac{-\downarrow}{2}}\!-
          \bar{\mbox{K}}^{\frac{+\uparrow}{2}}$ \\
         $\mbox{K}^{\frac{-\uparrow}{2}}\!-
          \bar{\mbox{K}}^{\frac{+\downarrow}{2}},
          \mbox{K}^{\frac{+\downarrow}{2}}\!-
          \bar{\mbox{K}}^{\frac{-\uparrow}{2}}$
         \end{tabular} &
         $1+2$ \\
         \noalign{\vskip 2mm}

         \begin{tabular}{c}
         kink with neither \\
         spin nor charge
         \end{tabular} &
         $\mbox{K}-\bar{\mbox{K}}$ &
         \begin{tabular}{c}
         $1+1$ \\
         $2+2$
         \end{tabular} \\
         \noalign{\vskip 2mm}

         polaron &
         \begin{tabular}{c}
         $\mbox{P}^{-\uparrow}\!-\mbox{P}^{+\downarrow},
          \mbox{P}^{+\downarrow}\!-\mbox{P}^{-\uparrow}$ \\
         $\mbox{P}^{-\downarrow}\!-\mbox{P}^{+\uparrow},
          \mbox{P}^{+\uparrow}\!-\mbox{P}^{-\downarrow}$
         \end{tabular} &
         \begin{tabular}{c}
         $4$ \\
         $\,4'$
         \end{tabular} \\
         \noalign{\vskip 2mm}

         \begin{tabular}{c}
         dip with either \\
         spin or charge
         \end{tabular} &
         \begin{tabular}{c}
         $\mbox{D}^{\uparrow}\!-\mbox{D}^{\downarrow},
          \mbox{D}^{\downarrow}\!-\mbox{D}^{\uparrow}$ \\
         $\mbox{D}^{+}\!-\mbox{D}^{-},\mbox{D}^{-}\!-\mbox{D}^{+}$
         \end{tabular} &
         $\,4+4'$ \\
         \noalign{\vskip 2mm}

         \begin{tabular}{c}
         dip with neither \\
         spin nor charge
         \end{tabular} &
         $\mbox{D}-\mbox{D}$ &
         \begin{tabular}{c}
         $4+4$ \\
         $\,4'+4'$
         \end{tabular} \\
         \noalign{\vskip 1mm}
         \end{tabular}
\end{ruledtabular}
\label{T:photoproduct}
\end{table}

   We close our time-dependent HF calculation estimating how probable each
photoproduct is.
We carry out numerical integrations with varying initial condition and
count in Fig. \ref{F:photoproduct} the resultant photoproducts classified
into several categories.
High-energy excitations above the electron-hole continuum tend to relax
into polarons, while excitons pumped within the optical gap are
self-localized and then dissociated into solitons and antisolitons in
pairs.
Solitons are easier neutral than charged.
Both solitons and polarons are so fragile that the major part of them
disappears in pairs within a few picoseconds.
There appear kinks and dips on the bond order parameters with unusual spin
and/or charge densities in their centers, which can be regarded as linear
combinations of prototypical solitons and polarons. \cite{H3835}
A pair of kink and antikink wearing half of the full electron/hole charge
and spin,
$\mbox{K}^{\frac{+\uparrow}{2}}\!-\bar{\mbox{K}}^{\frac{-\downarrow}{2}}$,
for example, consists of
$\mbox{S}^{0\uparrow}\!-\bar{\mbox{S}}^{0\downarrow}$ and
$\mbox{S}^{+}\!-\bar{\mbox{S}}^{-}$ in phase and of equal weight.
Such resonant states are better understandable if we observe
Figs. \ref{F:SSonCDW} and \ref{F:SSonACP} in more detail.
When a pair of kink and antikink separate away from each other or
recombine to disappear, not only do excess spin and charge densities
alternate, but even their signs are oscillating, in the centers of them.
Such striking fluctuations of spin and/or charge densities are
characteristic of photoproducts decaying nonradiatively rather than by
luminescence.

\section{Summary and Discussion}

\begin{figure}[b]
\centering
\includegraphics[width=86mm]{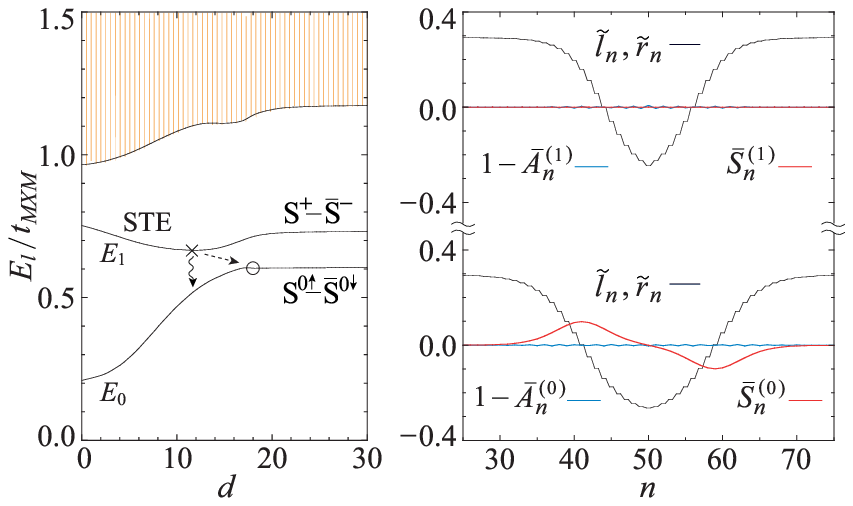}
\vspace*{-3mm}
\caption{%(Colour online)
         The solitonic relaxation channel of a photoexcited Pt$X$ chain
         whose ground state is of the CDW type, where
         $U_M/t_{M\!X\!M}=0.9$, $V_{M\!X\!M}/t_{M\!X\!M}=0.35$ and
         $\beta/\sqrt{t_{M\!X\!M}K_{M\!X}}=0.7$.
         Electronic structures of the lowest- and second-lowest-lying
         states, together with their background lattice configurations,
         are snapshotted at $d=18$ ($\circ$) and $d=12$ ($\times$),
         respectively.}
\label{F:ElPtCl}
\vspace*{6mm}
%\end{figure}

%\begin{figure}
\centering
\includegraphics[width=86mm]{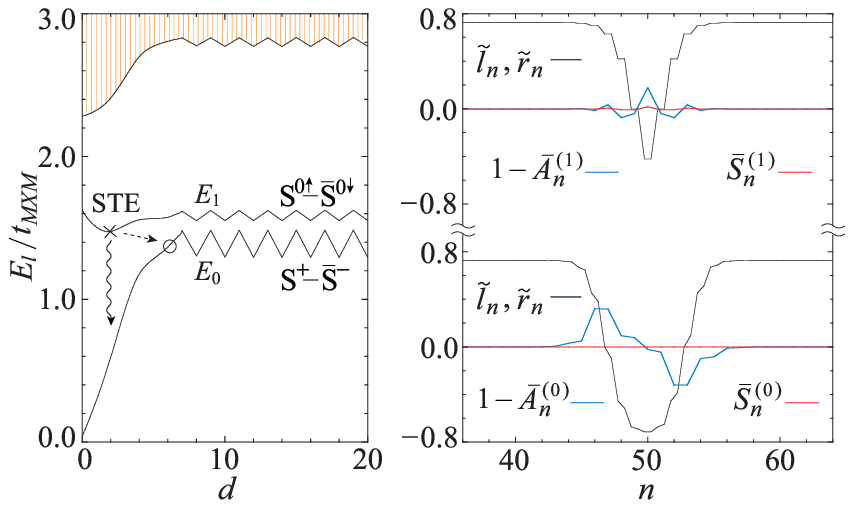}
\vspace*{-3mm}
\caption{%(Colour online)
         The same as Fig. \ref{F:ElPtI} but
         $U_M/t_{M\!X\!M}=0.5$, $V_{M\!X\!M}/t_{M\!X\!M}=0.05$ and
         $\beta/\sqrt{t_{M\!X\!M}K_{M\!X}}=1.0$.
         snapshots are taken at $d=2$ ($\circ$) and $d=6$ ($\times$).}
\label{F:ElPtI}
\end{figure}

   Coexistent neutral and charged solitons as stable photoproducts are
characteristic of Pt$_2X$ chains.
Mononuclear platinum complexes should be described with larger Coulomb
interactions, where neutral solitons are probable photoproducts, while
charged solitons are much less expected due to an energy barrier on the
way from an STE to them in a pair, as is shown in Fig. \ref{F:ElPtCl}.
Conversion of photoinduced charge-transfer excitons into neutral soliton
pairs has been indeed detected in the chloro and bromo complexes,
[Pt(en)$_2$Cl](ClO$_4$)$_2$ and [Pt(en)$_2$Br](ClO$_4$)$_2$.
\cite{S3066,K3049,K4245,O861}
The photoinduced valence anomalies are all ESR-active and a comparative
study of the heterometal compound
[Pt(en)$_2$][Pd(en)$_2$Br$_2$](ClO$_4$)$_4$,
whose ground state is no longer degenerate, enables us to distinguish
spin solitons from polarons.

   No doubt the simultaneous stabilization of neutral and charged solitons
against STEs is attributable to the reduction of Coulomb interactions,
but they owe much to the binuclear unit-assembled structure as well.
When we artificially model Pt$X$ chains with a similar set of correlation
and coupling parameters as adopted in the present Pt$_2X$ chains,
charged solitons indeed look available from STEs, but neutral solitons
turn hardly reachable instead, as is shown in Fig. \ref{F:ElPtI}.
Such a situation more or less agrees with experimental findings in PtI
chains.
There is a report \cite{O2248,O6330} that photocarriers in
[Pt(en)$_2$I](ClO$_4$)$_2$ may be charged solitons.
Since the halogen character increases in the platinum wave functions in
the order $\mbox{Cl}<\mbox{Br}<\mbox{I}$, the $d$-$p$ transfers are
enhanced and the Coulomb interactions are therefore suppressed effectively
in the iodo complexes.
The stabilization of charged solitons instead of neutral ones is thus
understandable.
The generated charged solitons are rather long-lived, possibly tunnelling
through a bumpy potential. \cite{O6330}

   The room-temperature conductivity of Pt$_2$(dta)$_4$I is about
nine orders of magnitude higher than those of typical Pt$X$ complexes.
\cite{K10068}
Photocarriers must be highly mobile in Pt$_2X$ chains.
Spin solitons, charged solitons and polarons are likely to come together
on such a fascinating stage.
The real-time simulation reveals quantum tunnelling between neutral and
charged solitons, breathing motion of charged solitons in a pair, and
asymmetry between electron and hole polarons, which are all stimulative
toward further experiments.

  The band filling of the dta family
 $M_2$($R$CS$_2$)$_4$I
($M=\mbox{Pt},\mbox{Ni}$; $R=\mbox{C}_n\mbox{H}_{2n+1}$)
\cite{B444,B2815,M11179,M2767}
might be varied in general because of their neutral chain structure.
Under doping, bipolarons, \cite{G6408,W6435,H5706,M5593,B6065}
that is, doubly charged bound polarons, also come into our interest.
Besides Pt$_2X$ chains, ladder-shaped Pt$X$ complexes have been recently
synthesized, \cite{K12066,K7372} exhibiting ground-state variations.
\cite{F044717,I063708,Y235116}
We hope our calculations will stimulate extensive optical explorations of
new varieties of metal-halide complexes.

%\acknowledgments
%
%   The authors thank S. Kuroda and K. Iwano for fruitful discussions.
%This work was supported by the Ministry of Education, Culture, Sports,
%Science, and Technology of Japan.

\end{document}